\documentclass[twocolumn]{aastex631}

\usepackage{newtxtext,newtxmath}
\usepackage[T1]{fontenc}

\usepackage{graphicx,graphics}	
\usepackage{color,epsfig}
\usepackage{amsmath}	
\usepackage{amssymb}	
\usepackage{psfrag}
\usepackage[T1]{fontenc}
\usepackage{ebgaramond}
\usepackage{newtxmath}
\usepackage{gensymb}
\usepackage{bm}

\newcommand{\Hi}{H\,\textsc{i}}
\newcommand{\HI}{H\,\textsc{i}~}

\def\HII{H\,\textsc{ii}~}

\def\bxHi{\bar{x}_{\rm H\,\textsc{i}}}
\def\Tb{{T_{\rm b}}}

\def\k{{\bm{k}}}
\def\a{{\bm{a}}}
\def\x{{\bm{x}}}
\def\r{{\bm{r}}}
\def\U{{\bm{U}}}

\def\cl{{\mathcal C}_{\ell}}

\def\n{\hat{\bm{n}}}
\def\v{\bm{v}}

\def\l{\mathcal L}

\def\dnu{\Delta \nu}
\def\nubar{\Bar{\nu}}

\def\mpcinv{{\rm Mpc}^{-1}}
\def\mpc{{\rm Mpc}}

\begin{document}


\title{Accurate parameter inference for the Light-cone Epoch of Reionization 21-cm signal}


\correspondingauthor{Suman Pramanick}
\email{suman21eor@gmail.com}

\author[0000-0002-3665-292X]{Suman Pramanick}
\affiliation{Department of Physics, Indian Institute of Technology Kharagpur, Kharagpur 721 302, India}

\author[0009-0006-0511-1991]{Anoop Krishna}
\affiliation{Department of Physics, National Institute of Technology Calicut, Calicut 673601, Kerala, India}

\author[0000-0001-7728-3756]{Rajesh Mondal}
\affiliation{Department of Physics, National Institute of Technology Calicut, Calicut 673601, Kerala, India}

\author[0000-0002-2350-3669]{Somnath Bharadwaj}
\affiliation{Department of Physics, Indian Institute of Technology Kharagpur, Kharagpur 721 302, India}

\begin{abstract}
The light-cone (LC) effect breaks the statistical homogeneity along the line-of-sight (LoS) direction of the 21-cm signal. Consequently, the traditional power spectrum (PS) fundamentally fails to capture the full two-point statistical information. The evolving power spectrum (ePS) offers an alternative, explicitly accounting for the LoS evolution due to the LC effect while maintaining physical interpretability. We compare the statistical constraining power of three different summary statistics: the standard cylindrical PS $P(k_\perp,k_\parallel)$, slice-wise PS $P_s(k, z)$ (1D-PS computed over small bandwidth LC slices), and ePS $P_e(k, z)$. We first demonstrate that $P_e(k,z)$ successfully recovers the benchmark 1D-PS of coeval simulations across all scales and redshifts, whereas the slice-wise PS only recovers the 1D-PS at small scales. To efficiently perform parameter inference, we train artificial neural network (ANN) emulators on $500$ simulated LC 21-cm signals. Our forecasts incorporate cosmic variance, estimated using $50$ statistically independent realizations of the signal, alongside SKA-Low system noise for integration times of $1000$ and $104$ hours. We demonstrate that the ePS outperforms its peers, yielding credible intervals that are $\sim 3$ and $\sim 1.4$ times tighter than those obtained using $P(k_\perp,k_\parallel)$ and $P_s(k,z)$, respectively. These results establish the ePS as an optimal summary statistic for interpreting forthcoming large-bandwidth LC 21-cm observations. 

\end{abstract}

\keywords{(cosmology:) dark ages, reionization, first stars - (cosmology:) large-scale structure of universe - (cosmology:)
diffuse radiation - methods: statistical – techniques: interferometric }



\section{Introduction}
\label{sec:intro}
The Epoch of Reionization (EoR) represents an important yet relatively unexplored frontier of modern cosmology. During this period, the first luminous objects of the Universe gradually heated and ionized the neutral hydrogen (\Hi) in the intergalactic medium (IGM).
Our current understanding of this epoch is primarily based on indirect observations, such as the Thomson scattering optical depth ($\tau_{\rm Th} = 0.054$; \citealt{collaboration2020planck}) of the cosmic microwave background (CMB), the Ly-$\alpha$ and Ly-$\beta$ Gunn-Peterson absorption troughs detected in the spectra of high redshift quasar \citep[{\it e.g.}~][]{Becker_2001,Fan_2003,fan2006constraining,gallerani2006constraining,becker2015evidence, Zhu_2021, Zhu_2022}, etc. Collectively, these probes constrain the epoch to the redshift range of $5.5\leq z\le 12$ \citep{Choudhury2022, Shaw_review}. Additionally, observations of high redshift Ly-$\alpha$ emitters \citep[{\it e.g.}~][]{ouchi2010statistics, hu2010atlas, ota2017new, Morales_2021}, Ly-$\alpha$ damping wings \citep[{\it e.g.} ][]{Banados_2018,Davies_2018,Wang_2020,Reiman_2020,yang2020measurements,greig2022}, the effective optical depth measurement of Ly-$\alpha$ and Ly-$\beta$ forest \citep[e.g.]{mcgreer2014model,Kulkarni_2019,Raste_2021,Zhu_2021,Zhu_2022} and recent JWST discoveries such as galaxies out $z \sim 13$ \citep[{\it e.g.} ][]{Witstok2025} continue to refine the constraints on EoR \citep{Chen_2026, Whitler2026, Mason2026}.

However, while these observations are extremely useful, they cannot provide a detailed, continuous spatial distribution and temporal evolution of the ionization process. The redshifted 21-cm emission line from \HI offers a direct probe that meets these requirements \citep{bharadwaj2005using, Furlanetto2006, Pritchard_2006, Zaroubi_2013}. However, detecting this faint signal is extremely challenging. The primary obstacle is the foregrounds that are $4-5$ orders of magnitude brighter than the expected signal \citep[e.g.][]{Ali_2008, Bernardi_2009, Bernardi_2010, Abhik_2012, Beardsley_2016}. Despite this severe observational challenge, significant progress has been made in placing increasingly stringent upper limits on the 21-cm power spectrum \citep{mertens2020improved} by several interferometers, including the GMRT\footnote{Giant Metrewave Radio Telescope; \url{http://www.gmrt.ncra.tifr.res.in}} \citep[][]{swarup1991, Paciga_2013}, PAPER\footnote{Precision Array for Probing the Epoch of Reionization; \url{http://eor.berkeley.edu/}} \citep[][]{Kolopanis_2019}, LOFAR\footnote{Low Frequency Array; \url{http://www.lofar.org}} \citep[][]{van2013lofar, yatawatta2013initial}, MWA\footnote{Murchison Widefield Array; \url{http://www.haystack.mit.edu/ast/arrays/mwa}} \citep[][]{bowman2013science, tingay2013murchison, dillon2014overcoming}, and HERA\footnote{Hydrogen Epoch of Reionization Array; \url{http://reionization.org}} \citep[][]{furlanetto2009cosmology, deboer2017hydrogen}. Currently, the best upper limit on the dimensionless power spectrum is $\Delta^2(k)=(21.4)^2 \, {\rm mK}^2$ at $z=7.9$ and $k=0.34\,h\,\mpcinv$ \citep{Abdurashidova_2023}, achieved by HERA. Additional constraints have been provided by LOFAR, yielding $\Delta^2(k=0.083\,\mpcinv,z=8.3)=(65.5)^2\,{\rm mK}^2$ \citep{lofar2025}, and by the MWA with $\Delta^2(k=0.156\,\mpcinv,z=8.2)=(98.67)^2\,{\rm mK}^2$ \citep{Trott_2025, sarkar2026}. Although a robust detection still has not been achieved, these increasingly stringent upper limits have already begun to place significant constraints on models of the EoR, ruling out broad regions of the astrophysical parameter space \citep[see e.g.,][]{Ghara_2020,Mondal_2020, Greig_2021, hera_2022, ghara2025}. The future looks promising with the Square Kilometre Array \citep[SKA;][]{mellema2013reionization, Koopmans_2015}, whose sensitivity is expected to enable not only precise measurements of the 21-cm PS but also tomographic imaging of the 21-cm signal.

To accurately interpret these current limits and prepare for future, accurate and fast models of the signal are essential. In addition to astrophysics, two observational effects, the redshift space distortions \citep[RSD;][]{bharadwaj2004cosmic, mao2012redshift,jensen2013probing,majumdar2013} and the light-cone effect \citep[LC;][]{barkana2006light,datta2012light, zawada2014light, datta2014light, mondal2018, pramanick2025quantifying}, play important roles in shaping the measured 21-cm signal. The LC effect, in particular, captures the continuous evolution of the reionization history, resulting in a statistically inhomogeneous signal along the line-of-sight~\citep[LoS;][]{mondal2019correction}. Consequently, the traditional Power Spectrum (PS) $P(k)$, which inherently assumes statistical homogeneity in all three directions, fails to capture the full two-point statistical information of the LC 21-cm signal \citep{trott2016exploring, mondal2018, Pramanick_2025, BlamartLiu2026}. Despite these facts, most parameter inference studies mentioned above use $P(k)$ as the summary statistic and ignore these effects due to computational complexity.

The multifrequency angular power spectrum (MAPS), $\cl(\nu_1,\nu_2)$ \citep{datta2007multifrequency}, bypasses the need for ergodicity along the LoS \citep{mondal2018} and has been proven to yield tighter constraints than the standard PS \citep{mondal2022param}. However, MAPS suffers from two practical drawbacks. First, the data volume is very large, scaling quadratically with the number of frequency channels for each angular multipole $\ell$. Second, its physical interpretation in terms of the length scales is not as straightforward as the one provided by the PS.
To bridge this gap, \cite{Pramanick_2025} recently introduced a new summary statistic, the evolving power spectrum (ePS) $P_e(k,z)$. This is designed to incorporate LoS evolution while maintaining the intuitive comoving length-scale interpretation of the standard PS.

In this work, we use an emulator-based framework to investigate the efficacy of the ePS for inferring parameters. We compare the performance of the ePS against two common alternatives, the standard cylindrical PS (cPS), $P(k_\perp,k_\parallel)$, and slice-wise PS (sPS), $ P_s (k,z)$. The latter is computed by dividing the full LC signal into narrow-bandwidth slices, thereby minimizing the LC effect. To ensure realistic comparisons, our MCMC forecasts incorporate both cosmic variance (estimated from an ensemble of $50$ statistically independent realizations of the 21-cm signal) and SKA-Low system noise.

This paper is structured as follows: Section~\ref{sec:LCsim} details our LC simulations and the parameters used in this study. In Section~\ref{sec:statistics}, we formally define the three summary statistics under consideration. Section~\ref{sec:emulator} outlines the emulators for different summary statistics. In Section~\ref{sec:noise}, we describe the method for incorporating the SKA-Low system noise. We present our main findings in Section~\ref{sec:results}, followed by a summary and discussion in Section~\ref{sec:summary}.

Throughout this paper, we adopt the Planck+WP best-fit values of the cosmological parameters $h=0.6704$, $\Omega_{\rm m0} = 0.3183$, $\Omega_{\Lambda 0} = 0.6817$, $\Omega_{\rm b 0} h^2 = 0.022 032$, $\sigma_8 = 0.8347$ and $n_{\rm s} = 0.9619$ from \citet{collaboration2020planck}.

\section{Simulating the light-cone 21-cm signal}
\label{sec:LCsim}
To generate the mock observational data, we simulate LC in a $(286.7\,{\mpc})^3$ comoving volume spanning from $z_f = 8.51$ at the far edge to $z_n = 7.53$ at the near edge, centered around $z_c=8.01$ (corresponding to a central frequency of $\nu_c=157.64\,{\rm MHz}$). This volume covers a comoving distance (frequency) from $r_f = 9300.14\,\mpc$ ($149.26\,{\rm MHz}$) to $r_n = 9013.42\,\mpc$ ($165.92\,{\rm MHz}$). Across this LoS extend, the mean \HI fraction varies from $[\bxHi]_f=0.65$ down to $[\bxHi]_n = 0.34$. The total frequency bandwidth $B=17.06\,{\rm MHz}$ is uniformly divided into $N_c = 512$ channels of width $\Delta\nu = 0.033\,{\rm MHz}$.

We construct the LC boxes by stitching together consecutive rectangular slices from an ordered series of coeval 3D snapshots along the LoS axis. Following this assembly, the LoS coordinates of the \HI particles are updated to account for RSD effects using the methodology described in \cite{mondal2018}. The coeval snapshots are generated using the established semi-numerical code {\sf ReionYuga}\footnote{\url{ https://github.com/rajeshmondal18/ReionYuga}} \citep{mondal2016, mondal2017}, which evolves the dark matter density field using $N$-body simulation and identifies dark matter halos using a Friends-of-Friends (FoF) halo finder \citep{mondal2015}. A comprehensive description of our specific LC simulation methodology can be found in \cite{pramanick2025quantifying}.

\begin{figure}
    \centering
    \includegraphics[width=\linewidth]{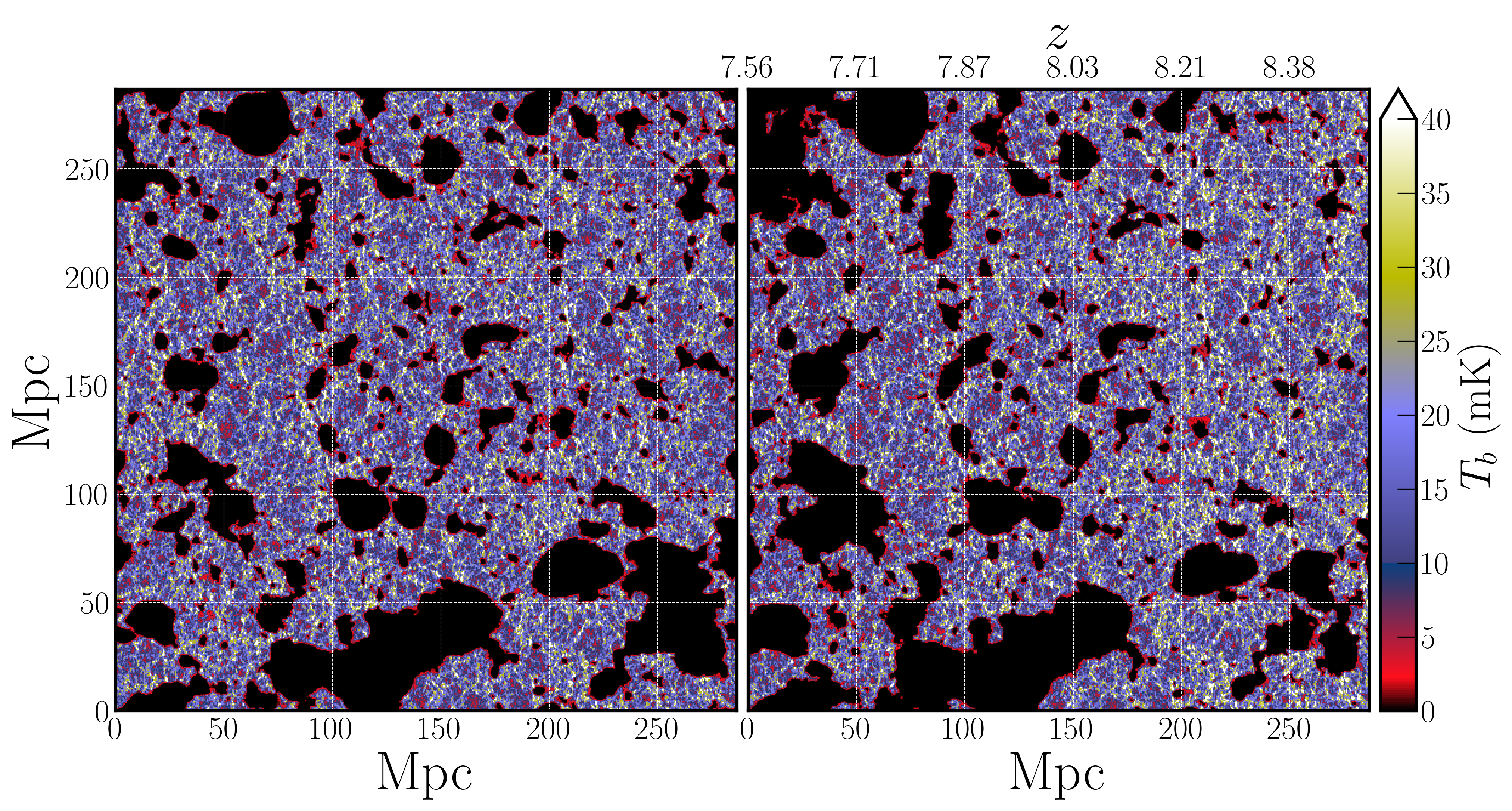}
    \caption{Two-dimensional slices of the 21-cm brightness temperature $\Tb$ maps for the coeval simulation (left) and the LC simulation (right), considering the fiducial parameter values. Both the simulations are at the central redshift $z_c=8$. For the LC simulation, the redshift $z$ corresponding to the LoS distance is shown on the upper x-axis.}
    \label{fig:tb}
\end{figure}

Our reionization model is governed by three parameters, $M_{\rm min}$, $N_{\rm ion}$ and $R_{\rm mfp}$. The $M_{\rm min}$ sets the minimum mass of the halos capable of hosting ionizing sources. $N_{\rm ion}$ represents the efficiency of ionizing photon production per baryon within these halos. Finally, $R_{\rm mfp}$ denotes the maximum comoving distance an ionizing photon can travel, which is analogous to the mean free path of the ionizing photon in the IGM. Physically, increasing $M_{\rm min}$ delays the onset of reionization by requiring more massive halos, which are rare at higher redshifts.
In addition, increasing $M_{\rm min}$ restricts the source of ionizing photons exclusively to high-density regions, resulting in an ionization field characterized by fewer but larger ionized regions. Whereas decreasing $M_{\rm min}$ allows small halos to host ionizing sources, thereby creating an ionization pattern with a large number of small ionized regions. Due to this, a change in $M_{\rm min}$ substantially changes the summary statistics we considered in our analysis, making it a sensitive parameter of our simulation \citep{shaw2019PS, Shaw2020}.
Increasing $N_{\rm ion}$ accelerates the reionization process. $R_{\rm mfp}$ has very little effect on most of the reionization history. It primarily influences the morphology of the ionized regions. The fiducial values for these parameters are set to $(M_{\rm min},\,N_{\rm ion},\,R_{\rm mfp}) = (12\times10^8\, {\rm M_\odot},\ 23.52,\ 16.95 \, {\rm Mpc})$. We refer the reader to \cite{shaw2020ps} for a detailed physical motivation of this specific parameterization. We have generated an ensemble of $50$ statistically independent LC realizations of this fiducial model to estimate the cosmic variance.

Figure~\ref{fig:tb} shows the brightness temperature $\Tb$ maps for the coeval simulation (left) and the LC simulation (right) considering the fiducial parameter values. The x-axis is the LoS axis where the redshift increases from left to right and reionization progresses from right to left. The same ionized \HII bubbles can be identified in both maps, with bubble sizes larger at $z<z_c$ and smaller at $z>z_c$ in the LC simulation compared to the coeval simulation. This visually represents that the statistical homogeneity present in the coeval simulation is broken due to the evolution of \HII bubble sizes in the LC simulation. 

\section{Summary statistics}
\label{sec:statistics}
The 21-cm brightness temperature fluctuations $\delta T_{\rm b} ({\bm r})$ can be conveniently expressed in terms of their Fourier conjugate $\Delta\tilde{\Tb}({\bm k})$ as
\begin{equation}\label{eq:tb}
\delta T_{\rm b} ({\bm r})=\int \frac{d^3 k}{(2 \pi )^3}e^{-i {\bm k}.r \hat{\bm n}} \Delta \Tilde{T}_b ({\bm k})\, .  
\end{equation}

\subsection{The Cylindrical Power Spectrum (cPS)}
If the underlying cosmological field is assumed to be statistically homogeneous, the traditional 1D power spectrum is defined as 
\begin{equation}\label{eq:ps1d}
\langle \Delta \Tilde{T}_b ({\bm k}) \Delta \Tilde{T}_b^{*} ({\bm k}') \rangle = (2 \pi)^3 \delta_{\rm D}^3 ({\bm k}-{\bm k}') P({\bm k})\, ,
\end{equation}
where $\delta_{\rm D}^3$ is the 3D Dirac delta function. However, the RSD and LC effects introduce anisotropy and non-ergodicity, respectively, along the LoS $({\bm \n})$. To isolate the LoS anisotropy, the 21-cm PS is usually decompose into $P({\bm k}_\perp,k_\parallel)$, where ${\bm k}_\perp$ and $k_\parallel$ are the perpendicular and parallel components of ${\bm \k}$ with respect to LoS direction ${\bm \n}$. However, $P({\bm k}_\perp,k_\parallel)$ still fundamentally relies on the assumption of statistical homogeneity and periodicity in all directions, which are not true along the LoS due to the LC effect. For our analysis, we estimate the bin-averaged $P(k_\perp,k_\parallel)$ by dividing the $({\bm k}_\perp,k_\parallel)$ space into $10$ equally spaced logarithmic bins along both axes.

\subsection{The Slice-wise Power Spectrum (sPS)}
One obvious approach to mitigate the LoS non-ergodicity is to analyze the signal over sufficiently smaller bandwidths where the evolution is minimal. To implement this, we divided the entire LC into thin slices along the frequency ($\nu$) axis. Specifically, the total bandwidth ($B$) is divided into $32$ equal slices, each with a width of $0.53$\,MHz. The transverse dimensions remain unchanged. We can approximate statistical homogeneity within these thin slices and independently estimate the standard $P({\bm k})$ for each slice using Equation~\ref{eq:ps1d}. We spherically bin the resulting ${\bm k}$-space into $10$ equally spaced logarithmic bins to estimate the bin-averaged sPS, $P_s(k,z)$. 

\begin{figure*}
    \centering
    \includegraphics[width=\linewidth]{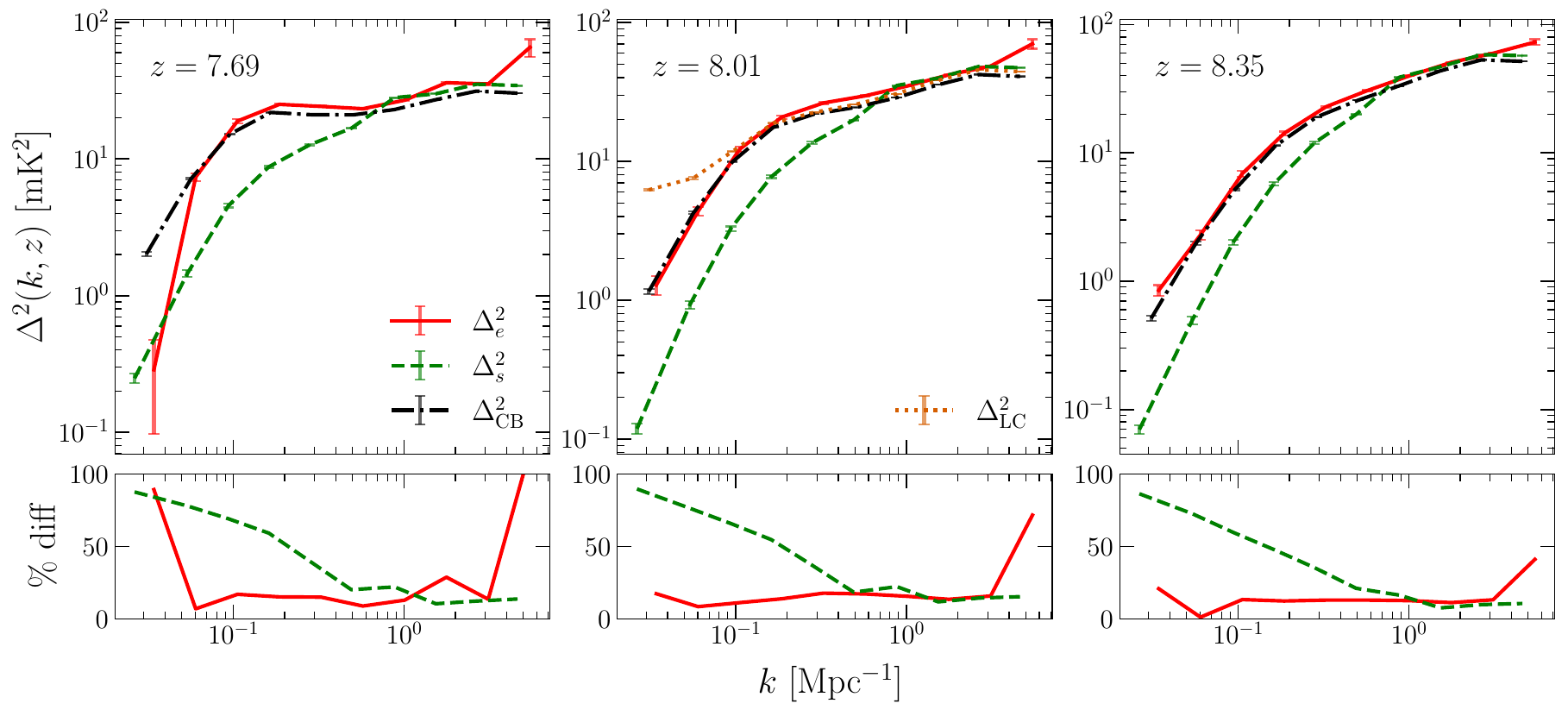}
    \caption{The left, middle, and right panels show different summary statistics at redshifts, $z=7.69,\, 8.01\, (z_c)\, {\rm and}\ 8.35$, respectively. The dot-dashed black lines show the dimensionless 1D-PS for the coeval simulation, $\Delta_{\rm CB}^2(k,z)$. The dotted yellow line in the middle panel shows the same quantity for the fiducial LC simulation. The dashed green and solid red lines show the spherically averaged dimensionless sPS $\Delta^2_s(k,z)$ and ePS $\Delta^2_e(k,z)$ for the fiducial LC simulation. The error bars represent $1\sigma$ cosmic variance errors. The bottom panels show the relative percentage differences for $\Delta^2_s$ (dashed green) and $\Delta^2_e$ (solid red) with respect to $\Delta_{\rm CB}^2$.}
    \label{fig:allps}
\end{figure*}

\subsection{The Evolving Power Spectrum (ePS)}
\label{sec:eps}
The ePS offers a physically motivated solution to the problem of analyzing the two-point statistics of non-ergodic LC 21-cm signal \citep{Pramanick_2025}. The ePS is formulated in close mathematical connection to the MAPS, $\cl(\nu_1,\nu_2)$. The 2D Fourier conjugate of the brightness temperature fluctuations on the sky-plane, $\delta T_{\rm b} (\bm{\theta},\nu)$, can be written as
\begin{equation}
    \Delta \tilde{T} ({\bm \ell}, \nu) = \int{d^2 \theta} e^{2\pi i {\bm \ell} \cdot   {\bm \theta}} \delta T_{\rm b}({\bm \theta}, \nu) \,.
\end{equation}
where, $\bm{\ell}$ is the Fourier conjugate of the angular coordinate ${\bm \theta}$. Under the flat-sky approximation, the MAPS is defined as 
\begin{equation}
    \cl (\nu_1,\nu_2) = \Omega^{-1} \left \langle \Delta \tilde{T}(\bm{ \ell},\nu_1) \Delta \tilde{T}^{*}({\bm \ell},\nu_2) \right \rangle\,,
\end{equation}
where the magnitude $\mid \bm{\ell} \mid = \ell$ corresponds to the angular multipole, and $\Omega$ is the solid angle subtended by the simulation volume at the position of the observer. Different $\ell$ values capture the homogeneous spatial clustering on the sky.

To estimate the ePS, we proceed by performing a variable change of the MAPS and express it as $\cl(\Delta \nu,\bar{\nu})$, where $\dnu = \nu_1 - \nu_2$ is the frequency separation and $\Bar{\nu} = (\nu_1+\nu_2)/2$ is the mean frequency. The $\dnu$ captures the statistically homogeneous part of the LoS spatial fluctuations. For an ergodic signal, the $\cl(\Delta \nu,\bar{\nu})$ would be invariant with respect to $\bar{\nu}$. Therefore, any residual dependence on $\bar{\nu}$ captures the LoS evolution. By mapping $\nubar$ to its corresponding $z$, the ePS is defined as \citep{Pramanick_2025}
\begin{equation}\label{eq:eps}
\cl(\Delta \nu,\bar{\nu}) = r_z^{-2} \, \int \frac{d k_\parallel}{(2 \pi)}  e^{-i  k_\parallel r_z' \Delta \nu}  P_e(k_\perp, k_\parallel,z) \,.  
\end{equation}
where $r_z$ is the comoving distance corresponding to $z$, $r_z' = dr_z/d\nu$, and,  $k_\parallel = {\bm k.\n}$ and $k_\perp = \ell/r_z$ are the components of ${\bm k}$ parallel and perpendicular to the LoS direction $\bm{\hat{n}}$. Note that, unlike the traditional PS (Equation~\ref{eq:ps1d}), this integral relation does not require statistical homogeneity along the LoS. Whereas it assumes statistical homogeneity along the new $\dnu$ axis, allowing it to capture the LoS evolution along $\nubar$ (or $z$). 

We keep the original channel resolution of $\cl(\dnu,\nubar)$ along the fast-varying $\dnu$ axis, along which the signal de-correlates rapidly. However, we have collapsed $16$ consecutive $\nubar$ to reduce
the number of channels along the slowly varying $\nubar$ or $z$ from $512$ to $32$. We further have discarded $4$ channels of $z$ (equivalent to $2.08\, {\rm MHz}$) from both near and far ends, as they are at the corners of $\cl(\dnu,\nubar)$, which does not contain sufficient $\dnu$ channels \citep[see Figure 2 of][]{Pramanick_2025}. This results in $24$ redshift bins of ePS for our LC simulation \citep{Pramanick_2025}. 

To compare ePS with the 1D-PS, we calculate the spherically averaged binned ePS by averaging the cylindrical ePS: 
\begin{equation}
\label{eq:pscy-mono}
P_e(k_\perp,k_\parallel,z)=P_e(k,z) \,.
\end{equation}
This operation averages out the LoS anisotropy. In practice, the $(k_\perp,k_\parallel)$ plane is divided into $10$ spherical logarithmic bins. The bin-averaged ePS $P_{e} (k,z)$ is then estimated from the measured $\cl(\Delta \nu,\bar{\nu})$ through a least-squares fitting using Equations~\ref{eq:eps} and \ref{eq:pscy-mono}. We refer the reader to \cite{Pramanick_2025} for a detailed derivation of the ePS.

\section{Comparison between the Summary Statistics}\label{sec:comparison}
Figure~\ref{fig:allps} shows a comparison of the three dimensionless ($\Delta^2(k,z) = k^3P(k,z)/2\pi^2$) spherically averaged summary statistics, 1D-PS, sPS, and ePS at three representative redshifts 7.69, 8.01 ($z_c$), and 8.35. The dot-dashed black lines show baseline 1D-PS $\Delta^2_{\rm CB}(k,z)$ computed from the full coeval simulation box at $z$. Because coeval simulations are statistically homogeneous in all directions, their 1D-PS represents the true statistical information of the field at that particular $z$. In the middle panel, we show the 1D-PS calculated from our fiducial LC simulation, $\Delta^2_{\rm LC}(k,z)$ (dotted yellow line), which differs from $\Delta^2_{\rm CB}(k,z)$ on large scales. The sPS $\Delta^2_{s}(k,z)$ (dashed green), mitigates this discrepancy to some extent, but it only follows baseline $\Delta^2_{\rm CB}(k,z)$ at smaller scales ($k \geq 0.8\,\mpcinv$).
The deviation of sPS at large scales mostly originates from a smaller number of large-scale modes along the LoS direction.
In contrast, the ePS $\Delta^2_{e}(k,z)$ (solid red), successfully recovers $\Delta^2_{\rm CB}(k,z)$ across almost all $k$ bins. The only noticeable deviations occur in the smallest $(0.03\,\mpcinv)$ and the largest $(5.47\,\mpcinv)$ bins. The excess noise in these $k$ bins for $\Delta^2_{e}(k,z)$ is due to the fact that they have much fewer modes. When spherically binning the $(k_\perp,k_\parallel)$ space to estimate $P_e(k,z)$ using Equation~\ref{eq:pscy-mono}, these two bins occupy the sparsely populated corners of the cylindrical grid, capturing significantly fewer independent modes and resulting in large sample variance. Thus, we discard these two extreme $k$-bins from all subsequent inference analysis. Percentage differences relative to $\Delta^2_{\rm CB}(k,z)$ for both $\Delta^2_{s}(k,z)$ and $\Delta^2_{e}(k,z)$ are shown in the lower panels. The error bars in the figure represent  $1\sigma$ cosmic variance errors estimated from 50 statistically independent realizations of the signal.

Crucially, the ability of the ePS to reliably recover the coeval 1D-PS across most scales and at all $z$ demonstrates its statistical strength to capture the LC effect-induced evolution. By successfully isolating this LoS statistically inhomogeneous information, the ePS proves to be a robust and superior summary statistic for parameter estimation. We explicitly quantify this ability in Section~\ref {sec:results}. We note here that ePS recovering coeval 1D-PS at each redshift may be a coincidence, and studying any deeper reason is beyond the scope of this paper. The fact that ePS accurately isolates the LoS evolution of an LC signal is also not necessarily a requirement for it to perform better than other summary statistics. Our claim here is purely a proof of concept, and we plan to investigate these in future works.




\begin{figure*}
    \centering
    \includegraphics[width=\linewidth]{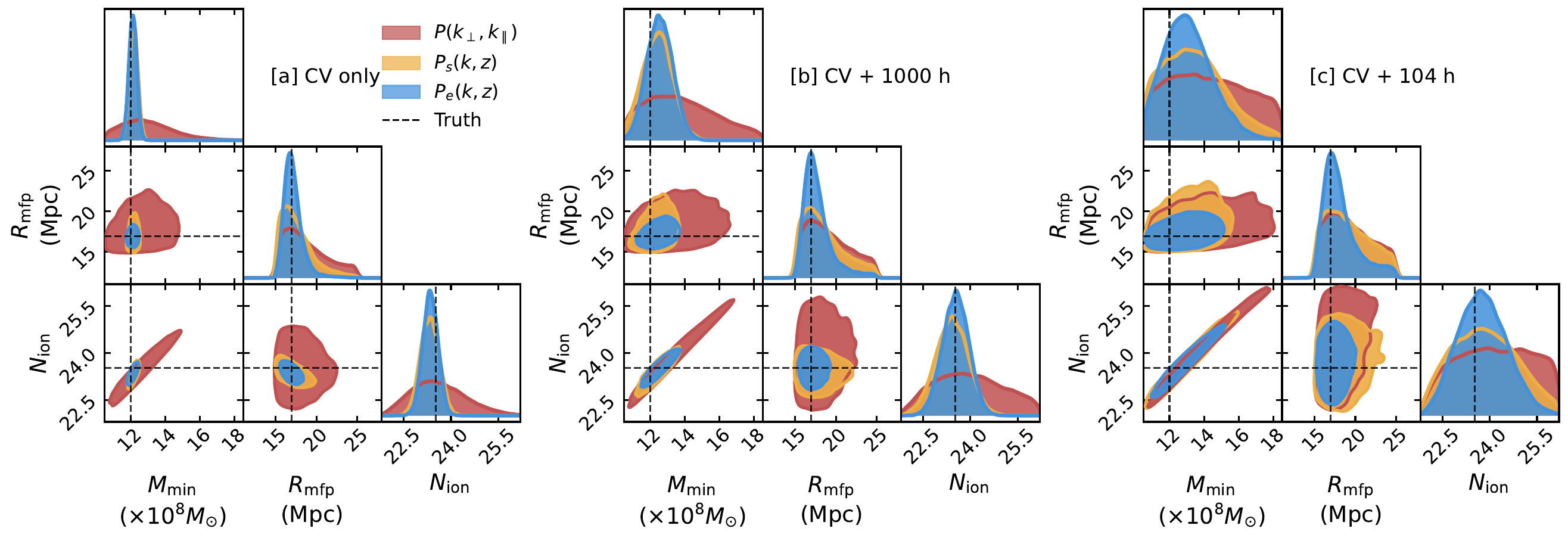}
    \caption{The 1D and 2D marginal $1\sigma$ posterior distributions for the three astrophysical parameters inferred using the cPS $P(k_\perp,k_\parallel)$, sPS $P_s(k,z)$, and ePS $P_e(k,z)$ summary statistics for the [a] CV only, [b] CV $+\,1000$ h and [c] CV $+\,104$ h cases. The red, orange, and blue contours show the constraints from $P(k_\perp,k_\parallel)$, $P_s(k,z)$, and $P_e(k,z)$, respectively. The dashed black lines indicate the fiducial parameter values.}
    \label{fig:1sigma}
\end{figure*}

\section{Emulators}
\label{sec:emulator}
Generating the LC 21-cm signal at every step of an MCMC chain using the methods described in Section~\ref{sec:LCsim} is computationally very expensive. We therefore accelerate our inference pipeline by emulating three summary statistics using artificial neural networks (ANNs). Our training and validation datasets are built upon 500 LC simulations. The parameters for these simulations are sampled from the 3D parameter space ($M_{\mathrm{min}}$, $N_{\mathrm{ion}}$, $R_{\mathrm{mfp}}$) using the Latin hypercube method. The sampled parameters span the following ranges: $M_{\mathrm{min}} (\times10^8\, {\rm M_\odot}) \in [10,\ 30]$, $N_{\mathrm{ion}} \in [16.21,\ 30.21]$, and $R_{\mathrm{mfp}}(\mathrm{Mpc}) \in [15,\ 25]$. From each LC output, we compute the cPS, ePS, and sPS, which serve as the training target. This dataset is randomly divided into training and test sets in a 70:30 ratio.

The bin-averaged cPS is calculated on a $10 \times 10$ grid in the $(k_\perp, k_\parallel)$ plane. For the ePS and sPS, we truncate the primary output grids from $32 \times 10$ bins to $24 \times 8$ bins by trimming the lowest and highest $k$ and $z$-bins, which (as discussed in Section~\ref{sec:comparison}) are dominated by sample variance. All three summary statistics are used in the dimensionless forms, with the two-dimensional arrays flattened into one-dimensional output vectors.

For each summary statistic, we constructed and trained a separate fully connected ANN to map the three input parameters to the corresponding high-dimensional statistical outputs. The network architectures are developed using the Keras package within the TensorFlow \citep{tensorflow2015-whitepaper,chollet2015keras}. All three emulators share a common architecture comprising five hidden layers with 256, 256, 128, 128, and 64 neurons. We use swish activation functions, $L_2$ weight regularization (with a penalty coefficient of $10^{-5}$), and a linear output layer with a size equal to the number of retained bins. The emulators achieve $R^2 > 0.99$ on the unseen test set, confirming that the networks capture more than $99\%$ of the variance in the data.

To quantify the residual modeling error, we follow the validation methodology established by \citet{Breitman2024}. We compute the fractional residual error in each retained output bin, introducing a noise floor in the denominator to avoid numerical divergences in bins where the signal is close to zero. We quantify this fractional-error distribution using its root-mean-square value, $F_{\mathrm{RMS}}$. This provides a more conservative estimate than the median statistic used by \citet{Breitman2024}, as the RMS is more sensitive to the tail of poorly predicted bins. The resulting modeling errors are found to be $F_{\mathrm{RMS}} = 2.97\%$, $4.28\%$, and $4.40\%$ for the cPS, sPS, and ePS emulators, respectively.


\section{The SKA-Low system noise}
\label{sec:noise}
To quantify the relative constraining power of these summary statistics under realistic observational conditions, we incorporate the anticipated SKA-Low system noise for integration times of $104$ and $1000$ hours in our forecasts. We consider an optimistic scenario in which the signal is contaminated only by the random Gaussian system noise and is free of foregrounds and other systematics.

The primary observables of radio interferometry are the visibilities recorded at baseline ${\bm U}={\bm d}/\lambda$ and at the corresponding observation frequency $\nu$, where ${\bm d}$ is the antenna separation projected onto the sky plane. We adopt the proposed SKA-Low baseline configuration from \citet{Dewdney2016}, consisting of $512$ stations, each with a diameter of $D = 35 \, {\rm m}$. We simulate a mock observation where the instrument tracks a target field at ${\rm Dec.} = -30 ^\circ$ for $8 \, {\rm h}$ per night. We generate continuous baseline tracks at various frequencies $(\nu)$ across the bandwidth of our simulation, following the comprehensive system noise simulation formalism detailed in \cite{shaw2019PS}.

The baselines are linearly mapped to the perpendicular component of ${\bm k}$ as ${\bm k_\perp} = 2\pi{\bm U}/r$, where $r$ is the comoving distance corresponding to $\nu$. We grid the ${\bm k_\perp}$ distribution onto the same grid used for our LC simulation. The baseline density is then calculated at each grid point ${\bm k_G}$ to estimate the baseline sampling function $\tau({\bm k_G})$ for each frequency slice. These $\tau({\bm k_G})$ is used in Equation~1 of \cite{shaw2019PS} to estimate the noise PS $P_N({\bm k_G})$ at each $\nu$ within the bandwidth of our observation. 

For each $\nu$, we generate system noise at every grid point ${\bm k_G}$ using 
\begin{equation}\label{eq:noise}
\Delta_N({\bm k_G}) = \sqrt{\frac{VP_N({\bm k_G})}{2}}\left[a({\bm k_G})+ib({\bm k_G}) \right],
\end{equation}
where $a({\bm k_G})$ and $b({\bm k_G})$ are Gaussian random variables with zero mean and unit variance, and $V$ is the survey volume. We stitch consecutive rectangular slices from these gridded noise maps along the LoS to construct the noise light cone. An ensemble of $50$ statistically independent realizations of these noise light cones was generated to directly estimate the system noise variance $\mathbf{\delta^N_X}$ for each of our summary statistics $(\rm{X})$. 

As mentioned in Section~\ref{sec:LCsim}, we use 50 statistically independent realizations of our fiducial LC to estimate the mean and cosmic variance $\mathbf{\delta^C_X}$ for each of our summary statistics. We have scaled the cosmic variance by a factor of $(1.73)^{-2}$ to account for the fact that the SKA-Low field of view $(\sim 3^{\circ})$ is approximately $\sim 1.73$ times the angular extent of our simulations $(1.79^{\circ})$. The total variance $\sigma_X^2$ for each statistic ${\rm X}$ is then taken as the sum of the scaled variance, the thermal noise variance, and the emulator-modeling variance $\delta^E_X$ as 
\begin{equation}
    \sigma_X^2 = \delta^C_X/(1.73)^{2} + \delta^N_X + \delta^E_X.
\end{equation}
The total error budget $\sigma_X$ is dominated by cosmic variance at large scales, and the system noise dominates at small scales. 


\section{Results}
\label{sec:results}
We execute our parameter inference using the affine-invariant MCMC ensemble sampler implemented in the \texttt{emcee}\footnote{\url{https://emcee.readthedocs.io}} Python package \citep{Foreman-Mackey2013}. We use the trained and validated emulators (Section~\ref{sec:emulator}) within the MCMC loop to rapidly generate model predictions for each step. We assume a standard diagonal Gaussian likelihood that incorporates the total error comprising cosmic variance, system noise, and modeling uncertainty. We explore the three-dimensional parameter space using 48 walkers, each evolved for 6000 steps. We assume uniform flat priors across the above-mentioned parameter bounds (Section~\ref {sec:emulator}). We carry out this inference independently for all three summary statistics for three specific observational scenarios: [a] CV-only limiting case, corresponding to infinite observation time;  [b] CV$+1000$\,h case; and [c] CV$+104$\,h case. The SKA-Low system noise (Section~\ref{sec:noise}) is incorporated into the latter two cases. 

\begin{figure*}
    \centering
    \includegraphics[width=\textwidth]{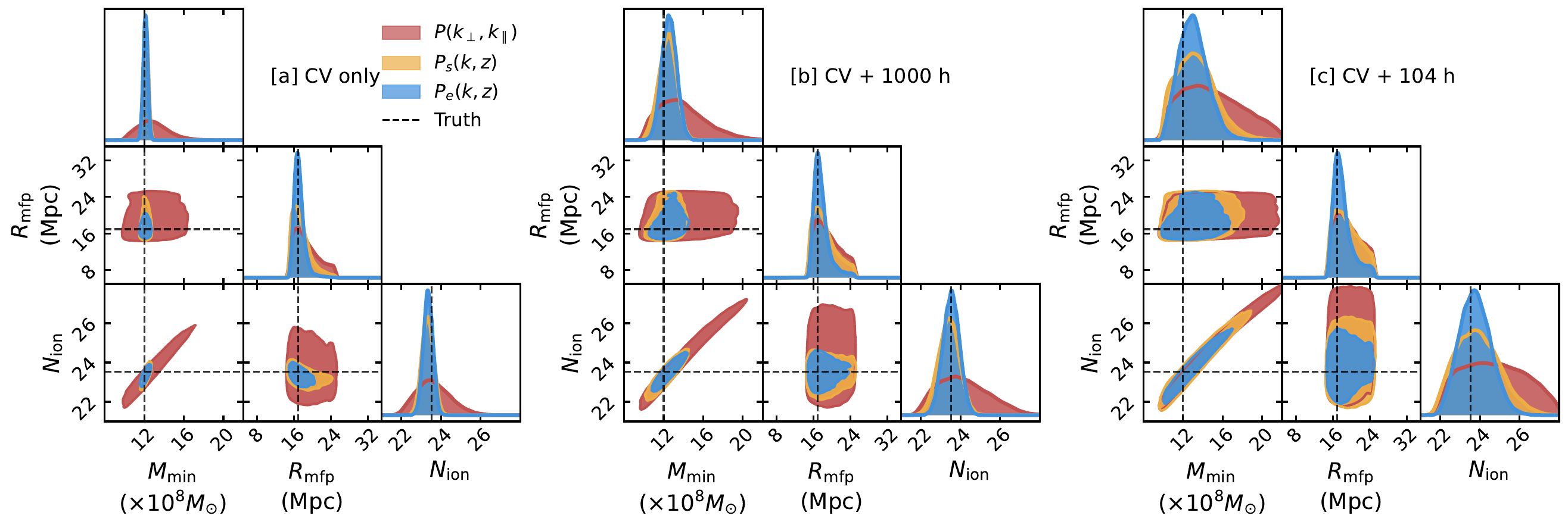}
    \caption{Same as Figure~\ref{fig:1sigma} but shows $2\sigma$ posterior distributions.
    }
    \label{fig:2sigma}
\end{figure*}

Figure~\ref{fig:1sigma} shows the 1D and 2D marginalized $1\sigma$ posteriors for the three specific observational scenarios considered here. The constraints derived from the cPS, sPS, and ePS are plotted in red, orange, and blue, respectively. The dashed lines indicate the true values. For all summary statistics, the posteriors are centered around the true values, confirming an unbiased parameter recovery. In Figure~\ref{fig:2sigma}, we expand this comparison. This figure presents the broader $2\sigma$ confidence contours for the same three observational scenarios.

Given this unbiased recovery, our primary focus is to evaluate the relative constraining power of each statistic, quantified by the widths of their respective credible intervals. We find that the ePS $P_{\rm e}(k,z)$ consistently yields the tightest parameter bounds, followed by the sPS $P_{\rm s}(k,z)$, with the standard cPS $P(k_\perp,k_\parallel)$ providing the weakest constraints. Averaged across the three parameters and three observational cases, the bounds generated by the ePS are roughly a factor of $3$ narrower than those obtained from the cPS, and approximately a factor of $1.4$ narrower than those derived using the sPS.

An earlier parameter estimation study by \cite{mondal2022param} showed that utilizing the full raw MAPS, $\cl(\nu_1,\nu_2)$, provides constraints roughly twice as tight as the traditional power spectrum. Later, \cite{Pramanick_2025} showed that despite having orders of magnitude less data volume than MAPS, the ePS $P_e(k,z)$ retains the bulk of the LoS information inherent to the MAPS. Our MCMC forecasts confirm this. We find that our parameter estimation using the ePS actually outperforms the previously explored MAPS implementations. The primary reasoning behind this superiority lies in strategic data compression. To reduce the overall data volume of the MAPS, \cite{mondal2022param} collapsed 8 consecutive frequency channels  (corresponding to $1.6$\,MHz). Along with the slowly varying $\nubar$ direction, this also collapses $\cl(\nu_1,\nu_2)$ along the fast varying $\dnu$ direction where the signal decorrelates mostly within $\sim 4$ MHz (see Figure~2 of \cite{Pramanick_2025}). This led to a loss of information. In contrast, the mathematical formulation of the ePS strategically collapses the transformed MAPS, $\cl(\dnu,\nubar)$, along the slowly evolving $\nubar$ axis, while retaining the high-resolution information along the $\dnu$ axis. Consequently, the ePS efficiently compresses the dataset without losing LoS information originally contained in the $\cl(\nu_1,\nu_2)$. This demonstrates that the ePS does more than just mimic a slice-wise analysis. It successfully extracts additional, unbiased information from the LC signal, making it a highly optimal statistic.


\section{Summary and Discussion} 
\label{sec:summary}
The LC effect is an unavoidable physical consequence of observing the EoR 21-cm signal over a large bandwidth, breaking the statistical homogeneity along the LoS. Consequently, the standard PS, which inherently assumes statistical homogeneity in all directions, cannot provide an accurate basis for quantifying the two-point statistics of the signal. While the MAPS $\cl(\nu_1,\nu_2)$ accounts for this LoS non-ergodicity and provides tighter constraints than the standard PS \citep{mondal2022param}, it suffers from two practical drawbacks. Its data volume scales quadratically with the number of frequency channels at each $\ell$, and it lacks the physical interpretation in terms of comoving length scales of the PS. The recently developed ePS \citep{Pramanick_2025} overcomes these issues by synthesizing the strengths of both summary statistics.

In this study, we comprehensively compared the parameter-constraining capabilities of three different summary statistics: the standard cylindrical PS $P(k_\perp,k_\parallel)$, slice-wise PS $P_s(k,z)$, and evolving PS $P_e(k,z)$. We used LC simulations spanning $17.06$\,MHz bandwidth, over which the mean \HI fraction varies from $0.65$ to $0.34$. While $P(k_\perp,k_\parallel)$ is sensitive to the anisotropies induced by redshift space distortion, it fundamentally relies on statistical homogeneity along the LoS. Therefore, it fails to quantify the full two-point statistical information in the LC signal. Conversely, $P_s(k,z)$ localizes this LoS evolution by dividing the full bandwidth into small ($0.53$ MHz) slices. However, this slice-wise approximation only manages to recover the true statistical information at small scales (Figure~\ref{fig:allps}). On the other hand, the $P_e(k,z)$ is successful in tracing the true information across nearly all scales (Figure~\ref{fig:allps}) at any given redshift. This confirms that ePS accurately captures the two-point information of an LC signal.

To translate these theoretical properties of the summary statistics into statistical constraining power, we evaluate parameter inference. We trained ANN-based emulators for the summary statistics using $500$ LC simulations. Using an affine-invariant MCMC sampler, we inferred three parameters of {\sf ReionYuga} framework ($M_{\rm min}$, $N_{\rm ion}$, and $R_{\rm mfp}$). Our forecasts incorporated cosmic variance and SKA-Low system noise for integration times of $1000$ and $104$\,hours. 

Note that the data volume of our summary statistics differs from each other. While for cPS the data volume is $10\times10$, for sPS and ePS it is $24\times 8$, respectively. However, in Figure~\ref{fig:collap} we compressed the data volume of sPS and ePS to $8\times 8$ by compressing $3$ consecutive redshift bins and find that our conclusions remain unchanged. 

Across every observational scenarios (Figures~\ref{fig:1sigma} and \ref{fig:2sigma}), $P_e(k,z)$ provides the tightest constraints. On average, the credible intervals produced by $P_e(k,z)$ were $\sim3$ and $\sim1.4$ times tighter than those produced by $P(k_\perp,k_\parallel)$ and $P_s(k,z)$, respectively. This improvement arises because the ePS makes more effective use of the redshift-dependent information in the LC signal, which also helps break parameter degeneracies typically encountered in the standard PS-based analyses. 

Our preliminary study is restricted to a particular reionization scenario using the {\sf ReionYuga} simulation and a particular reionization history and bandwidth. Therefore, the specific numbers quoted in this study may vary if a different reionization scenario or bandwidth is considered. However, the main conclusion of the work remains valid for any reionization scenario and bandwidth. We plan to study this in detail in our future work.
Furthermore, we have restricted this foundational analysis to the spherically averaged monopole of the ePS; the full cylindrical ePS $P_e(k_\perp,k_\parallel,z)$ retains an even richer information about LoS anisotropy in its higher-order multipole moments \citep{Pramanick_2025}. We plan to incorporate these higher-order components into our inference in future works, which promises to further enhance the parameter-constraining capabilities of next-generation 21-cm observations.

\section*{Acknowledgements}
The authors thank the anonymous reviewer for their thoughtful comments and suggestions. SP acknowledges support from the Prime Minister's Research Fellowship (PMRF). RM is supported by the NITC FRG Seed Grant (NITC/PRJ/PHY/2024-25/FR G/12). 
The authors acknowledge the super-computing facilities at the Centre for Theoretical Studies, Department of Physics, IIT Kharagpur.

\section*{Data Availability}
The simulated data and packages used in this work will be shared upon reasonable request to the authors.

\bibliography{main}{}
\bibliographystyle{aasjournal}




\appendix


\begin{figure}
    \centering
    \includegraphics[width=\linewidth]{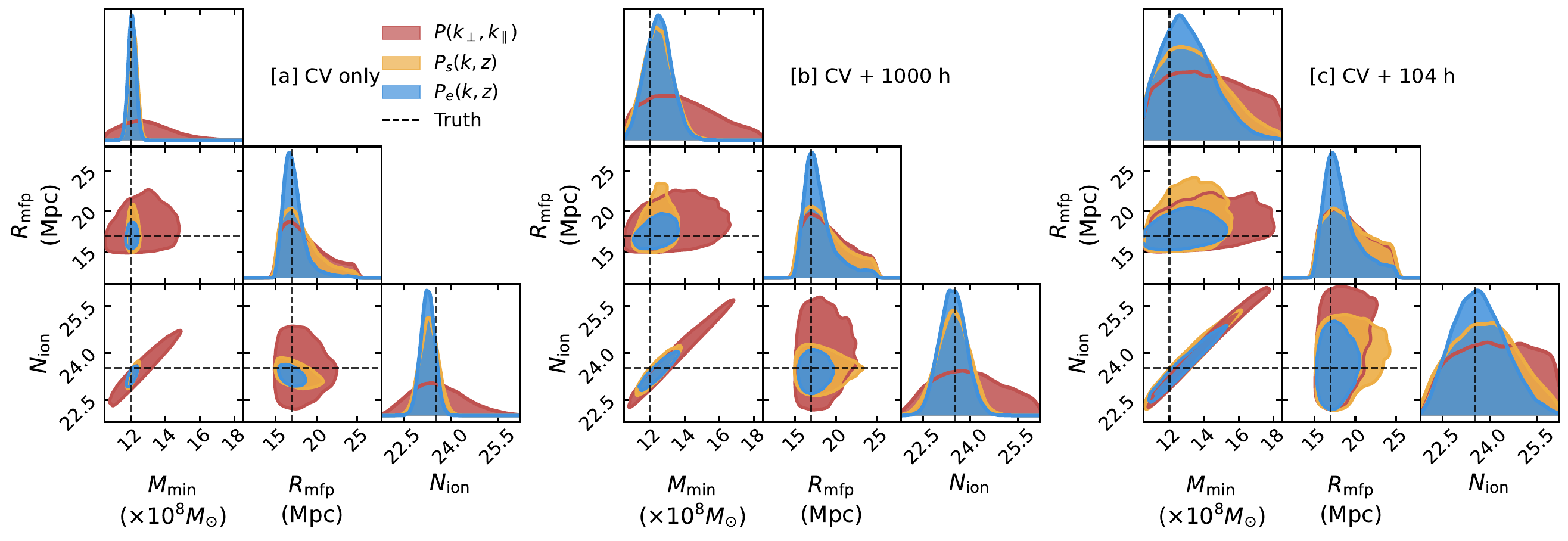}
    \caption{Marginalized posterior distributions and $1\sigma$ (68\%) credible contours obtained using original cPS and compressed sPS and ePS. We have averaged every three consecutive redshift bins of sPS and ePS to obtain an $8\times8$ data vector. We retain the original data volume $10\times10$ for the cPS. The ePS yields $2.74$ and $1.38$ times tighter constraints than cPS and sPS, respectively, which is close to the values obtained from Figure~\ref{fig:1sigma} that uses original data volume for all summary statistics. This suggests that, irrespective of data volume, ePS contains more information than sPS and cPS. } 
    \label{fig:collap}
\end{figure}
\label{lastpage}
\end{document}